\documentclass[preprint,aps,floatfix,showpacs]{revtex4}
\usepackage{dcolumn}
\usepackage{bm}
\usepackage{graphicx}
\begin{document}

\title{Dependence of nuclear binding on hadronic mass variation}
\author{V. V. Flambaum$^{1,3}$ and R. B. Wiringa$^2$}
\affiliation{
$^1$\mbox{Argonne Fellow, Physics Division, Argonne National Laboratory, 
          Argonne, IL 60439} \\
$^2$\mbox{Physics Division, Argonne National Laboratory, Argonne, IL 60439}
$^3$\mbox{School of Physics, University of New South Wales, Sydney 2052, 
          Australia} \\
}

\date{\today}

\begin{abstract}
We examine how the binding of light ($A\leq 8$) nuclei depends on
possible variations of hadronic masses, including meson, nucleon, and
nucleon-resonance masses.
Small variations in hadronic masses may have occurred over time; the 
present results can help evaluate the consequences for big bang
nucleosynthesis.
Larger variations may be relevant to current attempts to extrapolate
properties of nucleon-nucleon interactions from lattice QCD calculations.
Results are presented as derivatives of the energy with respect to
the different masses so they can be combined with different predictions 
of the hadronic mass-dependence on the underlying current-quark mass $m_q$.
As an example, we employ a particular set of relations obtained from a
study of hadron masses and sigma terms based on Dyson-Schwinger equations
and a Poincar\'{e}-covariant Faddeev equation for confined quarks and diquarks.
We find that nuclear binding decreases moderately rapidly as the quark mass 
increases, with the deuteron becoming unbound when the pion mass is increased
by $\sim$60\% (corresponding to an increase in $X_q=m_q/\Lambda_{QCD}$ of 2.5).
In the other direction, the dineutron becomes bound if the pion mass is 
decreased by $\sim$15\% (corresponding to a reduction of $X_q$ by $\sim$30\%).
If we interpret the disagreement between big bang nucleosynthesis
calculations and measurements to be the result of variation in $X_q$, we 
obtain an estimate  $\delta X_q/X_q=K \cdot (0.013 \pm 0.002)$ where $K \sim 1$
(the expected accuracy in $K$ is about a factor of 2). The result is dominated
by $^7$Li data.
\end{abstract}


\maketitle

\today

\section{Introduction}\label{introduction}

There are several reasons to search for a variation of the fundamental
``constants'' of nature in space and time. First, the Universe is evolving;
several phase transitions happened in the early Universe accompanied by 
dramatic changes in vacuum energy, fundamental masses,  
fundamental interactions (electromagnetic, weak, and strong)
and properties of elementary particles (e.g., confinement of quarks).
At later stages the equation of state of the Universe continued to evolve, 
from radiation domination (pressure $p= \epsilon/3$
where $\epsilon$ is the energy density), to cold-matter domination
($p \ll \epsilon$), and ``recently'', about 5 billion years ago,
to dark-energy domination ($p \approx - \epsilon$). 
In view of these dramatic changes it seems natural to check
if there is any evolution in the values of the fundamental
constants during this process.

A second reason is that spatial variation can
explain a fine-tuning of the fundamental constants which allows
humans (and any life) to appear.  Indeed, it is well-known that if the values
of some fundamental constants (e.g., related to the strong interaction)
would differ by even 1\% we could not appear. 
If we assume that the fundamental constants vary in space, this problem of 
fine-tuning may be resolved in a most natural way: we appeared in the area 
of the Universe where the values  of the fundamental constants are
consistent with our existence.

A third reason comes from theories unifying gravity and other interactions.
Some theories suggest the possibility of spatial
and temporal variation of physical ``constants'' in the Universe 
(see, e.g. \cite{Marciano,Uzan,Damour1,Damour:1994zq,Bekenstein,Wetterich,
Barrow,Olive}).  Moreover, there exists a
mechanism for making all coupling constants and masses of elementary
particles both space- and time-dependent, and influenced by local
circumstances.  The variation of coupling constants can be
non-monotonic (e.g., damped oscillations).

We can only measure the variation of dimensionless parameters which do not
depend on the units we use.
In the Standard Model the two most important dimensionless parameters are
the fine structure constant $\alpha=e^2/\hbar c$ and the ratio
of the electroweak unification scale (determined by the Higgs vacuum 
expectation value) to the quantum chromodynamics (QCD) scale 
$\Lambda_{QCD}$ (defined as the position of the Landau pole in
the logarithm for the running strong coupling constant,
$\alpha_s(r) \sim const/\ln{(\Lambda_{QCD} r/\hbar c)}$). 
The variation of the Higgs vacuum expectation value (VEV) leads to the 
variation of the electron mass $m_e$ and quark mass $m_q$ which are 
proportional to the Higgs VEV.
The present work considers effects produced by the variation of
$X_q=m_q/\Lambda_{QCD}$ where $m_q=(m_u+m_d)/2$ is the average 
current-quark mass.

Up to now a majority of publications about temporal and spatial variation
of the fundamental constants have considered effects of variation of 
$\alpha$.  However, the hypothetical unification of all interactions
implies that the variation of the strong interaction parameter
$X_q=m_q/\Lambda_{QCD}$ may be larger than the variation of the 
electromagnetic $\alpha$.
For example, the grand unification theories (GUTs) discussed in 
Ref.~\cite{Marciano} predict
\begin{equation}\label{eq:alpha}
 \frac{\delta X_q}{X_q} \sim 35\frac{\delta\alpha}{\alpha}
\end{equation}
The coefficient here is model-dependent but large values are generic
for grand unification models in which variations come from high energy
scales; they appear because the running strong-coupling constant and
Higgs constants (related to mass) run faster than $\alpha$.
Indeed, the electroweak (i=1,2) and strong (i=3) inverse coupling
constants have the following
dependence on the scale $\nu$ and normalization point $\nu_0$:
\begin{equation} \label{eqn_inv_alpha}
 \alpha_i^{-1}(\nu)=\alpha_i^{-1}(\nu_0)+b_i ln(\nu/\nu_0)
\end{equation}
In the Standard Model, $2\pi b_{i=1,2,3}=41/10, -19/6, -7$;
the electromagnetic $\alpha^{-1}=(5/3)\alpha_1^{-1}+\alpha_2^{-1}$ 
and the strong $\alpha_s=\alpha_3$.
In GUTs all coupling constants are equal at the unification scale,
$\alpha_i(\nu_0)\equiv\alpha_{GUT}$.
We may consider two possibilities: if we assume that
only $\alpha_{GUT}$ varies, then Eq.~(\ref{eqn_inv_alpha})
gives us the same shifts for all inverse couplings:
\begin{equation}
 \delta \alpha_1^{-1}=\delta \alpha_2^{-1}= \delta \alpha_3^{-1}=
 \delta \alpha_{GUT}^{-1} \ .
\end{equation}
We see that the variation of the strong interaction constant 
$\alpha_3(\nu)$ at low energy $\nu$ is much larger than the variation
of the electromagnetic constant $\alpha$, since 
$\delta \alpha_3/\alpha_3=(\alpha_3/\alpha_{1,2})\delta \alpha_{1,2}/
\alpha_{1,2}$ and $\alpha_3 \gg \alpha_{1,2}$.

The second possibility is the variation of the GUT scale ($\nu/\nu_0$
in Eq.~(\ref{eqn_inv_alpha})).  This gives
\begin{equation} 
 \delta \alpha_1^{-1}/b_1=\delta \alpha_2^{-1}/b_2= \delta \alpha_3^{-1}/b_3
\end{equation}
Note that in this case the variations have different signs since $b_1$ and
$b_{2,3}$ have different signs. However, we expect an even larger enhancement 
of the variation of 
$\alpha_3$ ($\delta \alpha_3/\alpha_3=(b_3 \alpha_3/b_{1,2} \alpha_{1,2})\delta \alpha_{1,2}/\alpha_{1,2}$).

The variation of $m/\Lambda_{QCD}$ can be estimated from the definition
of $\Lambda_{QCD}$. The running of $\alpha_s$ near the electroweak scale
is given by  
\begin{equation} \label{Lalphas}
 \alpha_s(\nu)^{-1} \approx b_s ln(\nu/\Lambda_{QCD})
\end{equation}
Let us take $\nu=m_z$ where $m_z$ is the $Z$-boson mass. The variation
of Eq. (\ref{Lalphas}) and relations above give 
\begin{equation}\label{Lalpha}
 \frac{\delta (m_z/\Lambda_{QCD})}{(m_z/\Lambda_{QCD})}=
 -\frac{1}{b_s\alpha_s(m_z)} \frac{\delta\alpha_s(m_z)}{\alpha_s(m_z)}
 \sim \frac{C}{\alpha(m_z)} \frac{\delta\alpha(m_z)}{\alpha(m_z)}
\end{equation}
The value of the constant $C$ here depends on the model used. However,
the enhancement $1/\alpha \sim 100$ should make the factor $C/\alpha$ large.
Note that $m_z$ (as well as $m_{e,q}$) is proportional to the Higgs VEV.  

If this  estimate is correct, the variation  in  $m_{e,q}/\Lambda_{QCD}$
may be easier to detect than the variation in $\alpha$. 
The cosmological variation of $m_q/\Lambda_{QCD}$ can be extracted from 
the big bang nucleosynthesis (BBN), quasar absorption spectra
and Oklo natural nuclear reactor data~\cite{FlambaumShuryak2002}.
For example, the factor of three disagreement between the calculations
and measurements of the BBN abundance of $^7$Li may, in principle, be 
explained by the variation of $m_q/\Lambda_{QCD}$ at the level of
$\sim 10^{-3}-10^{-2}$ \cite{DFW04} (see also recent work~\cite{BBN,CNOUV,KM}).
The claim of the variation of the fundamental constants based on the 
Oklo data in Ref.~\cite{Lam} is not confirmed by recent studies~\cite{Oklo}
which give a stringent limit on the possible variation of the resonance
in $^{150}$Sm during the last two billion years.
The search for the variation of $m_e/\Lambda_{QCD}$ using the quasar 
absorption spectra gave a non-zero result in Ref.~\cite{Ubach}
but zero results in Refs.~\cite{tzana,FK1}.
The present time variation of $m_{e,q}/\Lambda_{QCD}$
can be extracted from comparison of different atomic~\cite{tedesco}
and nuclear~\cite{th1,th4} clocks.
The review of the recent results on the variation of $\alpha$, 
$m_e/\Lambda_{QCD}$ and $m_q/\Lambda_{QCD}$ can be found in Ref.~\cite{F2007}. 

As we mentioned above, one can measure the variation of the dimensionless
parameter $X_q=m_q/\Lambda_{QCD}$.
Note that $m_q \approx $ 4 MeV $\ll \Lambda_{QCD} \approx $ 220 MeV.
As a result, nuclear parameters (nucleon mass, reaction cross-sections, etc.)
are determined mainly by $\Lambda_{QCD}$.  Therefore, in all calculations
it is convenient to assume that $\Lambda_{QCD}$ is constant and calculate
the dependence on the small parameter $m_q$. In other words, we measure
all masses in units of $\Lambda_{QCD}$ and will simply restore 
$\Lambda_{QCD}$ in the final results.
For example, the relation between the variation of the proton mass and quark
mass $\delta m_p/m_p= 0.064 \delta m_q/m_q$ should be understood as
$\delta X_p/X_p=0.064 \delta X_q/X_q $ where
$X_p=m_p/\Lambda_{QCD}$ and $X_q=m_q/\Lambda_{QCD}$.
This approach was formulated in Ref.~\cite{FlambaumShuryak2002}. 

Predictions for the relations between the variations of quark and hadron 
masses may come from a variety of approaches, such as chiral perturbation 
theory~\cite{BS03,EMG,FLTY,AALTY} or Dyson-Schwinger
equations (DSE)~\cite{FHJRW06}.
Nuclear binding energies and spectra would vary as a consequence
of the hadron mass variation, affecting a number of physical processes,
such as BBN and the Oklo phenomenon.
Large enough changes might alter the stability of some nuclei, e.g.,
unbind the deuteron, bind the dineutron, or even make $A$=5,8 nuclei 
particle-stable~\cite{WP02}.
In this work we estimate the changes in binding for $A$=2-8 nuclei that
result from small changes in hadronic masses.
We also evaluate the effect of much larger changes on the two-nucleon
systems, which may be relevant to present attempts to extrapolate
results from lattice QCD calculations~\cite{BBOS06}.
We first consider how changes in meson, nucleon, and nucleon-resonance 
masses would alter some representative nucleon-nucleon ($N\!N$) 
Hamiltonians.  We then solve for the energy of the two-body systems exactly
and calculate variational Monte Carlo (VMC) estimates for the larger nuclei
with these forces.
We report our results as derivatives of the energies with respect to
the different hadron mass changes, so that the results can be utilized
with different predictions of the coordinated changes between quark and
hadron masses.
Finally, we utilize the DSE predictions for hadron mass-dependence on the
quark mass as an example to explore the effects on nuclear spectra
and BBN.

\section{Nuclear Hamiltonian}

We examine several Hamiltonians of the form:
\begin{equation}
   H = \sum_{i} K_{i} + \sum_{i<j} v_{ij} + \sum_{i<j<k} V_{ijk} \ .
\label{eq:H}
\end{equation}
Here $K_i$ is a nonrelativistic kinetic energy operator, $v_{ij}$ is a
two-nucleon potential, and $V_{ijk}$ is a possible three-nucleon potential.  
We consider three different Argonne models for $v_{ij}$: Argonne~$v_{14}$
(AV14) and Argonne~$v_{28}$ (AV28) from 1984~\cite{WSA84}, and Argonne~$v_{18}$
(AV18) from 1995~\cite{WSS95}.
In conjunction with AV18, we will use the Urbana model IX (UIX) $V_{ijk}$
from 1995~\cite{PPCW95}.

The AV14 and AV18 models are conventional $N\!N$ potentials, while
the AV28 model has additional explicit $\Delta(1232)$ degrees of freedom.  
The AV14 and AV28 were constructed together and fit to the same $np$
phase-shift solution WI81 of Arndt and Roper~\cite{SAID} so they are 
phase-equivalent.  The intention was to use them in
parallel many-body calculations to study the effect of including explicit
$\Delta$'s in the nuclear Hamiltonian.
In practice, AV28 has proven difficult to use, so beyond the two-nucleon
system, only some triton~\cite{PRB92} and nuclear matter~\cite{W84} 
calculations have been reported.  
However, AV14 has been widely used in a variety of few-body~\cite{W91,PWP92} 
and dense nucleon matter~\cite{WFF88} calculations.
The AV18 is an updated version of AV14 containing charge-independence-breaking
(CIB) terms and a complete electromagnetic interaction.  
AV18 was fit directly to 4,301 $pp$ and $np$ data in the 1993 Nijmegen 
partial-wave analysis~\cite{Nijm93}, and AV18 and AV18+UIX have become 
standard Hamiltonians for {\it ab initio} calculations of light 
nuclei~\cite{PW01} and dense matter~\cite{AP97}.

All three of the Argonne potentials contain electromagnetic (EM) interaction,
long-range one-pion-exchange (OPE), intermediate-range attraction, 
and short-range repulsion, written as a sum of operator components:
\begin{equation}
   v_{ij} = v_{\gamma}(r_{ij}) + \sum_p [ v^p_{\pi}(r_{ij})
          + v^p_{I}(r_{ij}) + v^p_{S}(r_{ij}) ] O^p_{ij} \ .
\end{equation}
The number of operators $O^p_{ij}$ is 14, 28, or 18, as indicated by the
AVxx designation.  The different operators are discussed in detail below;
here we summarize the differences between the models.
The AV14 and AV28 potentials both use an average nucleon mass, while AV18 
keeps separate proton and neutron masses, which introduces a small 
charge-symmetry-breaking (CSB) term into $K_i$.
The $K_i$ for AV28 also depends explicitly on the $\Delta$ mass.
In AV14 and AV28, $v_{\gamma}$ is just the Coulomb interaction
between protons (with a form factor); in AV18 the magnetic moment
interaction, vacuum polarization, and other small EM terms are added.

The $N\!N$ part of OPE is the same in AV14 and AV28, and an average 
pion mass is used.  In AV18 there is a weaker OPE coupling strength, a 
slightly different form factor, and the small charge-dependent (CD) terms 
owing to the difference between neutral and charged pion masses are kept.
The intermediate-range attraction is due primarily to two-pion-exchange
(TPE) processes; in AV14 and AV18 this feature is approximated by using the 
square of the OPE tensor function $T(m_\pi r)$ as a phenomenological
radial shape for $v^p_I$ and adjusting strength parameters of the 14 or
18 operators to fit data.  In AV28, 14 operators with explicit 
$N\!N$--$N\Delta$--$\Delta\Delta$ couplings are added to those of AV14;
twelve of these have OPE range and two have intermediate and short range.
These produce much of the intermediate-range attraction explicitly through 
coupled-channel effects.  The coefficients of the first 14 $N\!N$ operators 
are refit to the data but are smaller in magnitude than in AV14.

The short-range repulsion may be attributed to the exchange of heavier
$\rho$ and $\omega$ mesons with suitable form factors for finite-size
effects, but in all the Argonne models it is treated phenomenologically. 
AV14 and AV28 use a Woods-Saxon radial shape with 14 strength parameters
while AV18 has a slightly more general shape with 26 parameters.

\subsection{Potentials}

The OPE potential between nucleons can be written as:
\begin{equation}
   v_{\pi}(r_{ij}) = f_{\pi NN}^{2} \left[ X_{ij} \tau_{i} \cdot \tau_{j}
                      + \tilde{X}_{ij} T_{ij} \right] \ .
\label{eq:vpi}
\end{equation}
Here $T_{ij} = 3\tau_{zi}\tau_{zj}-\tau_{i}\cdot\tau_{j}$ is the
CD isotensor operator which contributes when the difference
between neutral and charged pion masses is retained, and
\begin{eqnarray}
   && X_{ij} = \frac{1}{3} \left( X^{0}_{ij} + 2 X^{\pm}_{ij} \right) , \\
   && \tilde{X}_{ij} = \frac{1}{3} \left( X^{0}_{ij} - X^{\pm}_{ij}
                                      \right) , \\
   && X^{m}_{ij} =  \left(\frac{m}{m_{s}}\right)^{2} \frac{1}{3} mc^{2}
                \left[ Y(mr_{ij}) \sigma_{i} \cdot \sigma_{j} +
                           T(mr_{ij}) S_{ij} \right] \ .
\label{eq:xm}
\end{eqnarray}
The $S_{ij} = 3\sigma_{i} \cdot \hat{r}_{ij} \sigma_{j} \cdot \hat{r}_{ij}
- \sigma_{i} \cdot \sigma_{j}$ is the usual tensor operator and
$Y(mr)$ and $T(mr)$ are the normal Yukawa and tensor functions
\begin{eqnarray}
\label{eq:ypi}
   Y(mr)&=&\frac{e^{-\mu r}}{\mu r}\xi(r) \ , \\
\label{eq:tpi}
   T(mr)&=& \left(1+\frac{3}{\mu r}+\frac{3}{(\mu r)^{2}}\right) Y(mr)
            \xi(r) \ ,
\end{eqnarray}
where $\mu=mc/\hbar$ and a short-range form factor $\xi(r)$ has been 
incorporated that makes both $Y(mr)$ and $T(mr)$ vanish linearly at the origin.
In AV18, the $X^{0,\pm}$ are calculated with explicit $m_{\pi^0}$ and 
$m_{\pi^\pm}$ masses and the scaling mass $m_s \equiv m_{\pi^{\pm}}$.
In AV14 and AV28 an average $m_\pi = \frac{1}{3}(m_{\pi^{0}}+2m_{\pi^{\pm}})$
is used, so that $\tilde{X}_{ij}$ vanishes, and the scaling mass 
$m_s \equiv m_\pi$.
The coupling $f_{\pi NN}^{2}$=0.081 in AV14 and AV28, and 0.075 in the more 
modern AV18.

The intermediate-range and short-range terms in the potentials are given by
\begin{eqnarray}
\label{eq:vi}
   v^p_I(r_{ij}) &=& I^p T^2(m_\pi r) \ , \\
   v^p_S(r_{ij}) &=& (S^p + Q^pr + R^pr^2) W(r) \ ,
\label{eq:vs}
\end{eqnarray}
where $W(r)$ is a Woods-Saxon function.
For AV14 and AV28, $Q^p$=$R^p$=0, while in AV18, there are boundary
conditions on $v^p$ such that no more than two of the three $S^p$, $Q^p$, 
and $R^p$ are independent and free to be fitted for any given operator.  
The associated operators $O^p_{ij}$ include fourteen
charge-independent (CI) operators that are common to all the models:
\begin{eqnarray}
O^{p=1,14}_{ij} = [1, \sigma_{i}\cdot\sigma_{j}, S_{ij},
{\bf L\cdot S},{\bf L}^{2},{\bf L}^{2}(\sigma_{i}\cdot\sigma_{j}),
({\bf L\cdot S})^{2}]\otimes[1,\tau_{i}\cdot\tau_{j}] \ .
\label{eq:op14}
\end{eqnarray}
Here {\bf L} is the relative orbital angular momentum and {\bf S} is the
total spin of the pair.  The AV18 model has four additional CD and CSB terms:
\begin{equation}
   O^{p=15,18}_{ij} = [1, \sigma_{i}\cdot\sigma_{j},
S_{ij}]\otimes T_{ij} \ , \ (\tau_{zi}+\tau_{zj}) \ .
\end{equation}
These latter terms are small, but important for fitting the differences
between current $pp$ and $np$ scattering data.

There are fourteen additional operators in AV28 that involve explicit $\Delta$
degrees of freedom.  The first two are
\begin{eqnarray}
O^{p=15}_{ij} &=& (\sigma_{i}\cdot {\bf S}_j)(\tau_{i}\cdot {\bf T}_j)
               +  ({\bf S}_i \cdot\sigma_{j})({\bf T}_i \cdot\tau_{j})
               + H.c. \ , \\
O^{p=16}_{ij} &=& S^{II}_{ij}(\tau_{i}\cdot {\bf T}_j) 
               +  S^{II}_{ji}({\bf T}_i \cdot\tau_{j}) + H.c. \ ,
\end{eqnarray}
where ${\bf S}_i$ (${\bf T}_i$) is the transition spin (isospin) operator
for particle $i$ that changes a spin (isospin) $\frac{1}{2}$ state to a 
$\frac{3}{2}$ state.  The generalized tensor operator is
$S^{II}_{ij} = 3\sigma_{i} \cdot \hat{r}_{ij} {\bf S}_{j} \cdot \hat{r}_{ij}
- \sigma_{i} \cdot {\bf S}_{j}$ and H.c. denotes the Hermitian conjugate.
These operators are part of a generalized OPE contribution
\begin{equation}
   v^{II}_\pi(r_{ij}) = f_{\pi NN}f_{\pi N\Delta} 
            \left(\frac{m}{m_{s}}\right)^{2}  \frac{1}{3} mc^{2} 
            \left[Y(mr_{ij}) O^{15}_{ij} + T(mr_{ij}) O^{16}_{ij}\right]
\label{eq:vtrans}
\end{equation}
that produces $N\!N$--$N\Delta$ transitions.  The $\pi N\Delta$ vertex coupling 
is taken as $f_{\pi N\Delta} = 2f_{\pi NN}$.  Other operators provide OPE
contributions to \mbox{$N\!N$--$\Delta\Delta$}, \mbox{$N\Delta$--$\Delta\!N$},
\mbox{$N\Delta$--$N\Delta$}, \mbox{$N\Delta$--$\Delta\Delta$} and 
\mbox{$\Delta\Delta$--$\Delta\Delta$} transitions, with an additional coupling
$f_{\pi\Delta\Delta} = \frac{1}{5}f_{\pi NN}$ being introduced.  
Intermediate- and short-range interaction is also added in the diagonal 
\mbox{$N\Delta$--$N\Delta$} and \mbox{$\Delta\Delta$--$\Delta\Delta$} channels.
The full set of operators is detailed in Ref.~\cite{WSA84}.

The Urbana model of three-nucleon interaction (TNI) is written as a sum of
two-pion-exchange P-wave and remaining shorter-range phenomenological terms,
\begin{equation}
   V_{ijk} = V^{2\pi,P}_{ijk} + V^{R}_{ijk} ~.
\end{equation}
The structure of the TPE P-wave term is expressed simply as
\begin{equation}
   V^{2\pi,P}_{ijk} = \sum_{cyc} \left( A^P_{2\pi}
     \{X^m_{ij},X^m_{jk}\}
     \{\tau_{i}\cdot\tau_{j},\tau_{j}\cdot\tau_{k}\}
   + C^P_{2\pi} [X^m_{ij},X^m_{jk}]
     [\tau_{i}\cdot\tau_{j},\tau_{j}\cdot\tau_{k}] \right ) \ ,
\end{equation}
where $X^m_{ij}$ is the same spin-space function of Eq.~(\ref{eq:xm}) 
evaluated with the average pion mass, $\sum_{cyc}$ is a sum over the
three cyclic exchanges of nucleons $i$, $j$, and $k$, and 
$C^P_{2\pi}=\frac{1}{4}A^P_{2\pi}$.  The $A^P_{2\pi}$ and the
strength of the $V^{R}_{ijk}$ term are determined by fitting the 
binding energy of $^3$H~\cite{PPCW95} and the saturation density of 
nuclear matter~\cite{AP97} in the presence of a given $v_{ij}$;
the parameters of UIX were selected in conjunction with AV18.

\subsection{Dependence on hadron masses}

We consider changes in the Hamiltonian that would be induced by
small changes in the hadron masses.  However, we do not consider changes
in the coupling constants that might occur due to mass-dependent loop
corrections~\cite{BS03,EMG}.  Variations of the nucleon
and nucleon-resonance masses alter the kinetic energy operator:
\begin{equation}
   K_i = -\frac{\hbar^2}{2m_i}\nabla^2_i + (m_i-m_N)c^2 \ ,
\label{eq:ki}
\end{equation}
where $m_i$ is the average nucleon mass $m_N=\frac{1}{2}(m_p+m_n)$ for AV14, 
but may be $m_\Delta$ part of the time for AV28.  For AV18 the kinetic energy 
operator has no $m_\Delta$ component, but does include a small CSB piece:
\begin{equation}
   K_i = -\frac{\hbar^2}{2m_N}\nabla^2_i
-\frac{\hbar^2}{2m_N}\left(\frac{m_n-m_p}{m_n+m_p}\right)\tau_{zi}\nabla^2_i \ ,
\end{equation}
where $m_N = 2m_pm_n/(m_p+m_n)$.
There is also a small dependence on the nucleon mass in the magnetic 
moment parts of the EM interaction in AV18.

Changes in $m_N$ and $m_\Delta$ also affect the energy expectation value
for AV28 through effects of the $N\!N$--$N\Delta$--$\Delta\Delta$ 
coupled channels.  This can be visualized in a simple closure approximation 
(see Eq.~(33) of Ref.~\cite{WSA84} and the accompanying discussion)
for the TPE diagrams where one or two intermediate $\Delta$'s are excited, 
propagate, and then de-excited by the transition potentials:
\begin{eqnarray}
\label{eq:trans}
  V_{\rm eff} \approx 
  && 2 V_{N\!N-N\Delta} \frac{-1}{\bar{E}_1+(m_\Delta-m_N)} 
       V_{N\Delta-N\!N} \\ \nonumber
  &+&  V_{N\!N-\Delta\Delta} \frac{-1}{\bar{E}_2+2(m_\Delta-m_N)} 
       V_{\Delta\Delta-N\!N} \ .
\end{eqnarray}
The two mean-energy denominators $\bar{E}_1$ and $\bar{E}_2$ would vary,
in part, as the kinetic energies of the intermediate states:
\begin{eqnarray}
  \bar{E}_1 &\approx& \frac{\hbar^2\bar{k}^2}{2m_N}
                    + \frac{\hbar^2\bar{k}^2}{2m_\Delta} \ , \\
  \bar{E}_2 &\approx& 2\frac{\hbar^2\bar{k}^2}{2m_\Delta} \ ,
\label{eq:tke}
\end{eqnarray}
with $\bar{k}$ an average intermediate momentum.
This physical effect can be approximated in the AV14 and AV18 potentials by 
multiplying the intermediate-range strength coefficients $I^p$ of
Eq.~(\ref{eq:vi}) by factors $(1+\delta_N)$ and $(1+\delta_\Delta)$.
The above equations represent only some of the terms contributing to the 
$m_N$ and $m_\Delta$ dependence introduced by the coupled channels, so we 
will fix the $\delta_N$ and $\delta_\Delta$ terms for AV14 by requiring that 
calculated two-body energies have the same mass dependence as AV28 without 
these factors.  We will take the same factors as approximately correct
modifications for AV18 also.

The dependence on the pion mass (which vanishes in the chiral limit $m_q=0$) 
can be obtained from the original pion-exchange interaction
which is proportional to the second derivative of the exponential potential,
$\nabla_i \nabla_j [\exp{(-\mu r)}/r]$ --- see, e.g.~\cite{FlambaumShuryak2002}.
As a result, the variation of the pion mass affects the strength of the
OPE through the $m^3$ dependence of $X^m_{ij}$ in Eq.~(\ref{eq:xm}) and 
through the range of the Yukawa functions $Y(mr)$ and $T(mr)$, 
Eqs.~(\ref{eq:ypi}-\ref{eq:tpi}).
Note that the scaling mass $m_s$ is not allowed to vary (this is only 
a coefficient which does not vanish in the chiral limit $m_q=0$). 
The dependence of OPE terms on pion mass is thus the same as given
in Eq.~(10) of Ref.~\cite{BS03}.
Neglecting the small effect of our short-range form factor $\xi(r)$, the
function $m^3 Y(mr)$ has a constant volume integral.  If $m$ is increased,
$m^3 Y(mr)$ will be larger inside $r=2/\mu$ and smaller outside.
However, $m^3 T(mr)$ will be smaller at all values of $r>0$.

The generalized OPE of AV28 that couples to intermediate $N\Delta$ and 
$\Delta\Delta$ states effectively incorporates considerable
multi-pion exchange effects.  We can approximate this real physical effect 
in AV14 and AV18 by assuming the intermediate-range potential has the
mass-dependence $v^p_I(r) \propto [m^3_\pi T(m_\pi r)]^2$.  
The closure approximation of Eq.~(\ref{eq:trans}) that justifies this 
connection applies most clearly to the first six static ({\bf L}-independent) 
operators $O^p_{ij}$ of Eq.~(\ref{eq:op14}) because of the closed
algebra of the spin (isospin) and transition-spin (isospin) operators.
However, the {\bf L}-dependent operators can also be affected through
the coupled-channels, so we will also consider changing these terms
in the intermediate-range potentials.  We will also consider changes 
to the residual $v^p_I(r)$ terms in AV28.  Thus we will study variations 
attributable to the pion mass in four stages with our Hamiltonians: 
(1) change in the OPE part of the potential $v_\pi$ (and $v^{II}_\pi$,
etc. for AV28) only; 
(2) change in both $v_\pi$ and the static TPE parts $v^{p=1,6}_I$ (TPE-s); 
(3) these changes plus the {\bf L}-dependent TPE parts $v^{p=7,14}_I$ (TPE-L); 
and (4) these changes in $v_{ij}$ plus the TPE part of the three-nucleon 
potential $V^{2\pi,P}_{ijk}$ (TNI).

Representing the dependence on heavier meson masses 
$m_\rho \approx m_\omega \approx m_V$ is much more problematic 
because the Argonne models do not have explicit heavy-meson exchange.
However, the effect of a short-range interaction
is determined mostly by the volume integral of this interaction, e.g.,
the volume integral gives us the strength constant $B$
if we want to approximate a short-range interaction by a zero-range
interaction $B \delta(\bf{r})$.
Therefore, we may approximate the dependence on masses of heavier mesons
by a change in the range parameters $r_0$ and $a$ of the short-range
Woods-Saxon potential
\begin{equation}
   W(r) = \frac{1}{1+{\rm exp}[(r-r_0)/a]} \ .
\end{equation}
used in Eq.~(\ref{eq:vs}).  To keep the same relative
variation of the volume integral for the Woods-Saxon 
potential and the meson exchange potential $\exp({-m_V r})/r$
 we change the parameters at the rate:
\begin{equation}
   \frac{\delta r_0}{r_0} = 
   \frac{\delta a}{a}     = -\frac{2}{3}\frac{\delta m_V}{m_V} \ .
\end{equation}
In this case, as $m_V$ increases, the range of the Woods-Saxon decreases
while the value at the origin remains constant and the volume integral
varies as $m^{-2}_V$.

The repulsive core of the three-nucleon potential $V^{R}_{ijk}$ may be
attributed to a combination of multi-pion and heavy-meson exchanges,
and also some relativistic effects~\cite{FPF95}.  Owing to its complicated
nature and phenomenological treatment and the fact that it gives a small 
contribution to energy expectation values, we do not attempt to determine 
its variation with changes in the hadronic masses.

\section{Energy Calculations}

We calculate the energies of the $^1$S$_0(np)$ virtual bound state and
the deuteron for AV28, AV14, and AV18 by direct solution of the two-body
equations.  The energies of $^3$H, $^{3,4,5}$He, $^{6,7}$Li, and $^{7,8}$Be 
are calculated for AV14 and AV18+UIX using variational Monte Carlo (VMC) 
methods.  
The VMC method is described in detail in Ref.~\cite{PW01} and references
therein.  Briefly, we construct suitably parametrized trial wave functions
$\Psi_V$ and evaluate the upper bound to the ground-state energy,
\begin{equation}
   E_V = \frac{\langle \Psi_V | H | \Psi_V \rangle}
              {\langle \Psi_V   |   \Psi_V \rangle} \geq E_0 \ ,
\label{eq:expect}
\end{equation}
using Monte Carlo techniques for the multi-dimensional integration.
The parameters in $\Psi_V$ are varied to minimize $E_V$, and the lowest value
is taken as the approximate solution.
We use a variational wave function of the form
\begin{equation}
   |\Psi_V\rangle = \left[1 + \sum_{i<j<k}U^{TNI}_{ijk} \right]
     \left[ {\cal S}\prod_{i<j}(1+U_{ij}) \right] |\Psi_J\rangle \ .
\label{eq:psiv}
\end{equation}
The $U_{ij}$ and $U^{TNI}_{ijk}$ are noncommuting two- and three-nucleon 
correlation operators, reflecting the spatial and operator dependence of 
$v_{ij}$ and $V_{ijk}$, and ${\cal S}$ is a symmetrization operator.
The form of the antisymmetric Jastrow wave function $\Psi_J$ depends on 
the nuclear state under investigation.
For the s-shell nuclei the simple form
\begin{equation}
   |\Psi_J\rangle = \prod_{i<j}f_c(r_{ij}) |\Phi_A(JMTT_{3})\rangle
\end{equation}
is used.  Here the $f_c(r_{ij})$ are central (spin-isospin 
independent) two-body correlation functions and $\Phi_A$ is an 
antisymmetrized spin-isospin state, e.g.,
\begin{equation}
   |\Phi_{4}(0 0 0 0) \rangle
        = {\cal A} |\uparrow p \downarrow p \uparrow n \downarrow n \rangle \ ,
\end{equation}
with ${\cal A}$ the antisymmetrization operator.  The $\Psi_J$ for p-shell
nuclei is more complicated; details are given in Ref.~\cite{PW01}.
The two-body correlation operator $U_{ij}$ is a sum of spin, isospin, 
and tensor terms:
\begin{equation}
   U_{ij} = \sum_{p=2,6} u_p(r_{ij}) O^p_{ij} \ ,
\end{equation}
where the $O^p_{ij}$ are the static operators of Eq.~(\ref{eq:op14}).
The central $f_c(r)$ and noncentral $u_p(r)$ pair correlation functions
are generated by a set of six coupled, Schr\"{o}dinger-like, differential
equations which include the $v_{ij}$ and a number of embedded variational
parameters.  These parameters are optimized in the energy minimization
and then kept fixed as the hadronic masses are varied.  The correlations 
are adjusted, however, because the altered interaction is used to
regenerate the correlations in each case.

The energies we obtain are shown in Table~\ref{tab:energy} and compared
to experiment.  The VMC method is reasonably accurate for $s$-shell nuclei, 
giving energies $\sim$2-3\% less bound than exact methods like Faddeev or 
Green's function Monte Carlo for a given Hamiltonian~\cite{PW01}.  
It is not as good for $p$-shell binding energies, but many other features 
such as density distributions and electromagnetic moments are in good
agreement.  We anticipate that small changes in the binding energies 
induced by small changes in hadron masses will be accurately tracked with 
the VMC calculations.
The comparison between AV14 and AV18+UIX models indicates the importance of 
including a three-nucleon interaction to approach the experimental energies.

We evaluate the mass-dependence of the energies of the two- and multi-nucleon 
systems by changing the hadron masses $m_H$ one at a time, increasing the
value by 0.1\% and calculating the resulting energy, and then decreasing by 
0.1\% and repeating the energy calculation.  The VMC calculations
follow the same random walk stored from the initial energy evaluation
to reduce the contribution of statistical noise.
Results given in the following three tables are the dimensionless derivatives 
of the energy with respect to changes in the hadron masses
\begin{equation}
 \Delta\mathcal{E}(m_H) = \frac{\delta E/E}{\delta m_H/m_H} \ .
\end{equation}
These results can then be combined with any given model for how the
different hadron masses are correlated with the underlying quark masses
to give a total binding energy prediction:
\begin{equation}
 E(m_q) = E(m_{q_0}) \left[ 1 + \sum_{m_H} \Delta\mathcal{E}(m_H) 
          \frac{\delta m_H(m_q)}{m_H} \right] \ ,
\label{eq:energy}
\end{equation}
where $m_{q_0}$ is the physical current-quark mass.  A specific example will be
given below.

\subsection{Two-nucleon energies}

The two-nucleon results for $\Delta\mathcal{E}(m_H)$ are given in 
Table~\ref{tab:two} for all three Hamiltonians.  
A simple approximate relation between changes in the deuteron binding 
energy $Q$ and virtual bound state energy $\epsilon_v$ is given 
by~\cite{DFW04}:
\begin{equation}
     \frac{\Delta\epsilon_v(m_H)}{\Delta Q(m_H)}
      \approx -\frac{\sqrt{Q}}{\sqrt{\epsilon_v}} \ .
\label{eq:vbs}
\end{equation}
Despite the wide range of values for $\Delta\mathcal{E}(m_H)$ in 
Table~\ref{tab:two}, this relation is valid within a factor of $\sim$2 for 
fifteen out of seventeen pairs of results.  The two exceptions are for the
OPE components of $\Delta \epsilon_v(m_\pi)$ for AV14 and AV18, where the 
sign is incorrect.  This discrepancy will be discussed below.

Changing the nucleon mass in AV14 and AV18 primarily changes just the 
kinetic energy component and is very similar for the two models: a 
larger $m_N$ translates to smaller $\langle K \rangle$ and more binding 
and thus a positive derivative $\Delta Q(m_N)$ for the deuteron.  
For the virtual bound state, the effect is the same but greater attraction 
corresponds to a reduction in the magnitude of $\epsilon_v$ and a
negative $\Delta \epsilon_v(m_N)$.

Changing $m_N$ in AV28 produces a larger change in the energies due to
the additional coupled-channel effects of Eqs.~(\ref{eq:trans}-\ref{eq:tke}) 
discussed above.  The change is more dramatic for $\epsilon_v$ than 
for $Q$, probably because the $^1$S$_0$ $N\!N$ channel can couple to
both $N\Delta$ and $\Delta\Delta$ intermediate states, while the 
deuteron can have only $N\!N$ and $\Delta\Delta$ components due to
isospin conservation.  To approximately incorporate this physical
effect into the phase-equivalent AV14 model, we can multiply its 
intermediate-range strength parameters $I^p$ of Eq.~(\ref{eq:vi}) by a 
factor $(1+\delta_N)$.  We choose a value $\delta_N = 0.49 \delta m_N/m_N$
that matches the mass dependence in the deuteron as shown by the line 
in Table~\ref{tab:two} labeled ``$m_N + \delta_N$''.  
This same factor approximately fixes the virtual bound state also.
Using the same $\delta_N$ factor in AV18 produces a change about one third 
larger.  This difference is probably due to the fact that AV18 is fit to a
more recent data set, with a weaker OPE coupling accompanied by
more-attractive intermediate-range terms and correspondingly 
more-repulsive short-range terms.  

Increasing the $\Delta$ mass in AV28 reduces the binding energy both 
through the one-body term of Eq.~(\ref{eq:ki}) and through the
coupled-channel effects of Eq.~(\ref{eq:trans}).  
Thus $\Delta\mathcal{E}(m_\Delta)$ has a sign opposite to
$\Delta\mathcal{E}(m_N)$.  To approximate this affect for AV14, we
can again multiply its intermediate-range strength parameters $I^p$
by a factor $(1+\delta_\Delta)$.  We take
$\delta_\Delta = -0.57 \delta m_\Delta/m_\Delta$ which gives a fair
reproduction of the behavior of AV28 for both the deuteron and 
the virtual bound state as shown by the line in Table~\ref{tab:two} 
labeled ``$\delta_\Delta$''.  Using the same factor in AV18 again
produces a larger rate of change.

We note that if the nucleon and $\Delta$ masses change at about the
same rate, from whatever the underlying quark mass dependence, 
then the effect on the AV28 energies could be obtained by the sum of 
the $m_N$ and $m_\Delta$ terms in Table~\ref{tab:two}, which is not 
very different from the $m_N$ term alone for AV14 and AV18 or from
the sum of $m_N + \delta_N$ and $\delta_\Delta$ terms.  In fact, these
corresponding sums of terms agree within 10-20\% for all three models.

Increasing the pion mass reduces the regularized OPE tensor function 
$m_\pi^3 T(m_\pi r)$ at all values of $r$.  
The binding of the deuteron is primarily due 
to the tensor coupling, so $Q$ is reduced and $\Delta Q$ 
is negative for all the models as shown by the line in Table~\ref{tab:two} 
labeled ``$m_\pi$~(OPE)''.  The value for AV28 is larger because of
the multi-pion-exchange effects included through the generalized OPE
potential.  The AV14 and AV18 values are smaller and close together, the 
difference between them being proportional to the different $f^2_{\pi N\!N}$
coupling constants used in the two models.  

Increasing $m_\pi$ in the $^1$S$_0$ channel, which depends only on 
$m_\pi^3 Y(m_\pi r)$, has the effect of slightly increasing the attraction
in the virtual bound state, making $\Delta(\epsilon_v)$ negative for the 
AV14 and AV18 models.  However, in the AV28 model, the generalized OPE provides 
significant intermediate-range attraction $\propto [m^3_\pi T(m_\pi r)]^2$, 
which is enough to reverse the sign of $\Delta(\epsilon_v)$ to be positive.
As the AV14 and AV18 cases here are the only two out of seventeen in 
Table~\ref{tab:two} that violate the relation between $\Delta Q$ and
$\Delta(\epsilon_v)$ of Eq.~(\ref{eq:vbs}), it appears that changing
only the OPE part of a conventional $N\!N$ potential like AV14 or AV18
is an incomplete representation of the physics in the singlet channel.

Changing $m_\pi$ in the static TPE part of the interaction has a more dramatic
effect than changing just the OPE part, as shown in Table~\ref{tab:two}
by the line labeled ``$m_\pi$~(+TPE-s).''  The $\Delta Q$ increases by 
by a factor of 2.5--4 in the deuteron for the AV14 and AV18 models compared 
to OPE only, while for AV28 it increases by a smaller factor of 1.5.  More
dramatically, the AV14 and AV18 values for $\Delta \epsilon_v$ change
sign and come into agreement with Eq.~(\ref{eq:vbs}).  The further 
addition of the non-static ``$m_\pi$~(+TPE-L)'' terms makes no difference
to the virtual bound state, as these operators do not contribute in the
$^1$S$_0$ channel, and rather small changes in the deuteron.

Finally, increasing the heavy-meson mass $m_V$ reduces the range of
the Woods-Saxon repulsion and increases the binding of
the deuteron, so $\Delta Q$ is positive.  The reduced repulsion for
the virtual bound state correspondingly makes $\Delta \epsilon_v$
negative.  In both the full-pion and heavy-meson exchanges, the AV18 has
larger $\Delta\mathcal{E}$ values than AV14 or AV28.  Again, this is
probably due to the weaker OPE and compensatingly larger intermediate-range
attraction and short-range repulsion.

\subsection{Multi-nucleon energies}

The multi-nucleon results for $\Delta\mathcal{E}(m_H)$ are given in
Table~\ref{tab:av14} for AV14 and Table~\ref{tab:av18uix} for AV18+UIX.
For every nucleus, and for every $m_H$ component, the signs are the
same as for the deuteron.  The relative sizes of the terms are also
about the same as the deuteron, with the exception of the $m_\pi$~(OPE)
term.  Because the light nuclei have approximately equal numbers of 
deuteron-like and $^1$S$_0$-like pairs~\cite{W06}, the 
$\Delta\mathcal{E}(m_H)$ are expected to be averages of the two sets
of trends in Table~\ref{tab:two}.  The anomalous behavior of 
$\Delta \epsilon_v$ in the $m_\pi$~(OPE) case discussed above causes 
these numbers to be much smaller in the multi-nucleon systems, and thus 
out of proportion compared to all the other terms.  As for the
deuteron, the multi-nucleon terms have a more rapid dependence with
AV18+UIX than with AV14.  However, the explicit $m_\pi$ contribution from
the three-nucleon force is very small.

\section{Dependence of nuclear binding energies and Big Bang nucleosynthesis
 on quark mass}

As an example of how to incorporate our nuclear binding energy results 
with a specific prediction for hadronic mass variation, we utilize the
results of a Dyson-Schwinger equation study of sigma terms in light-quark
hadrons~\cite{FHJRW06}.  Equations (85-86) of that work gives the rate of 
hadron mass variation as a function of the average light current-quark mass 
$m_q = (m_u+m_d)/2$ as:
\begin{equation}
 \frac{\delta m_H}{m_H} 
  = \frac{\sigma_H}{m_H} \frac{\delta m_q}{m_q}
\label{eq:dse}
\end{equation}
with $\sigma_H/m_H$ values of 0.498 for the pion, 0.030 for the $\rho$-meson,
0.043 for the $\omega$-meson, 0.064 for the nucleon, and 0.041 for the 
$\Delta$.
The values for the  $\rho$ and  $\omega$-mesons were reduced to 0.021 and 
0.034, respectively, in a subsequent study~\cite{HMRW}.  
We will use an average of the $\rho$ and $\omega$ terms
of 0.03 for the variation of our short-range mass parameter $m_V$.

\begin{figure}
\includegraphics[angle=-90,width=6.25in]{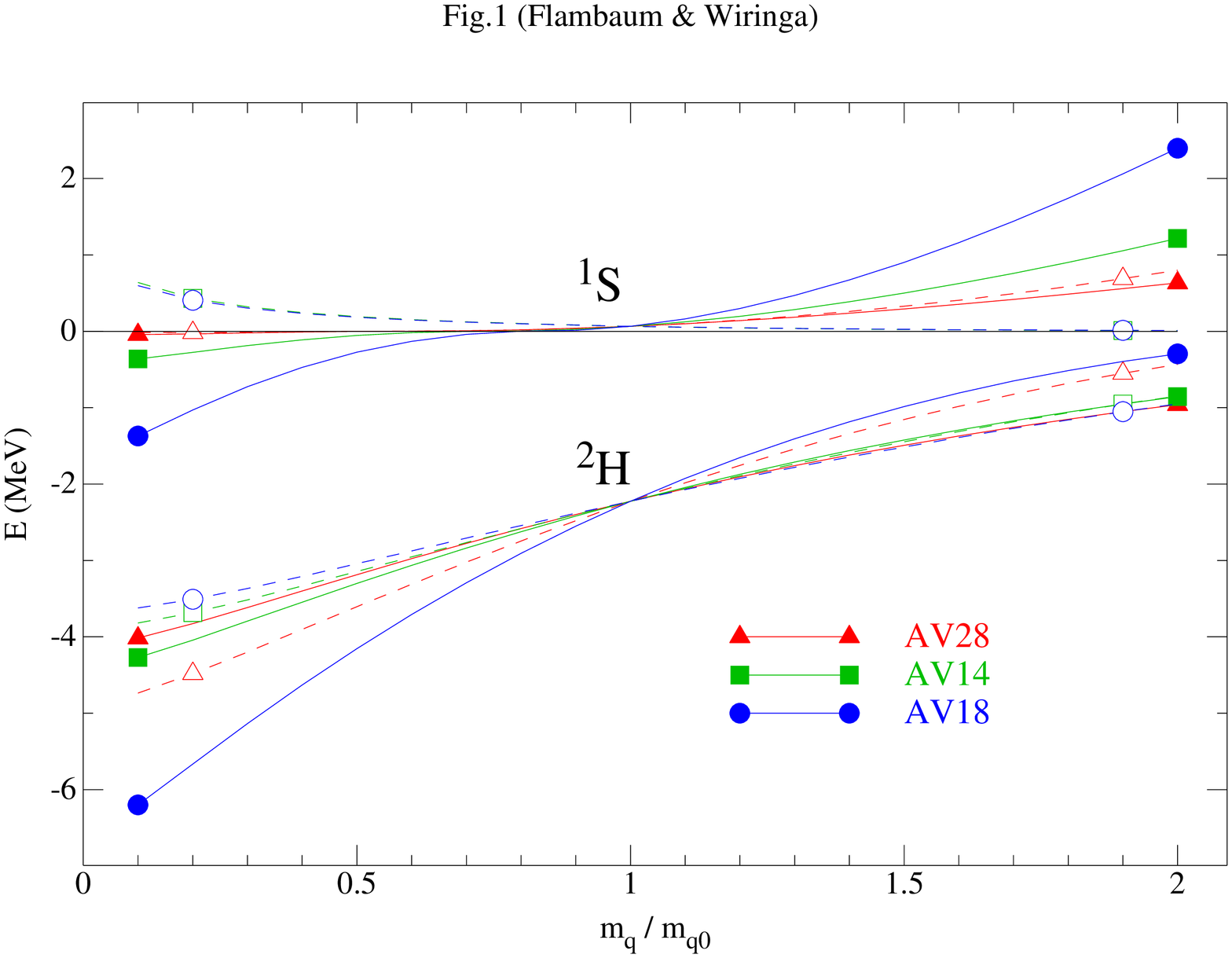}
\caption{(Color online) Variation of two-nucleon energies with current-quark
mass variation from DSE calculation: full calculation (solid lines) and with
OPE modification only (dashed lines) for three different Argonne Hamiltonians.
Virtual bound state energies are plotted as positive quantities;
$m_{q0}$ is the physical current-quark mass.}
\label{fig:two}
\end{figure}

In Fig.~\ref{fig:two} we show direct calculations for the two-nucleon states,
i.e., the deuteron and $^1$S$_0$($np$) states, for a range of $m_q$ values for
all three Hamiltonians.  The virtual bound state energies are plotted as a
positive quantity; when the $^1$S$_0$ energy is negative it indicates a true
bound state.  The dashed lines show the results from changing only the
pion mass in the OPE part of the interaction, corresponding to the line
$m_\pi$~(OPE) in Table~\ref{tab:two}.  To evaluate over this wide range
of $m_q$, we use the Gell-Mann-Oakes-Renner (GMOR) relation 
$m^2_\pi \propto m_q$.

The solid lines show the most complete calculation for each Hamiltonian. 
For the AV28 model this is the sum of the terms $m_N$, $m_\Delta$, 
$m_\pi$~(+TPE-L), and $m_V$, using GMOR for the pion and
the DSE values above for the variation of all other $m_H$.
For the AV14 and AV18 models, it is the sum of the terms $m_N + \delta_N$, 
$\delta_\Delta$, $m_\pi$~(+TPE-L), and $m_V$.  
As discussed in Sec.III.A, the $\delta_N$ and $\delta_\Delta$ effects 
are incorporated by multiplying the intermediate-range
strength parameters $I^p$ of Eq.~(\ref{eq:vi}) by a factor
\begin{equation}
 (1+\delta_N)(1+\delta_\Delta) 
  = (1+.49\frac{\delta m_N}{m_N})(1-.57\frac{\delta m_\Delta}{m_\Delta})
\end{equation}
which is $\approx (1+.008~\delta m_q/m_q)$ for the DSE values above.

At the OPE level, the AV14 and AV18 models show almost exactly the same
behavior for the $^1$S$_0$ state, with a gradually increasing attraction as
$m_q$ increases, while for the deuteron they both show a more rapid decrease
in binding, consistent with the results obtained in Refs.\cite{BS03,EMG}.  
If these trends continue, the deuteron will eventually move 
above the singlet state somewhere in the range 3-4 $m_q$.
The AV28 model has a somewhat more rapid dependence for the deuteron but 
its singlet state parallels the deuteron, becoming less attractive for 
larger $m_q$.  
In the other direction, the singlet state becomes a true bound state
at $\approx 0.7 m_q$, but it always remains above the deuteron.  
This different behavior is a consequence of the multi-pion exchange that 
is built into the AV28 model through the generalized OPE coupling to 
intermediate $\Delta$'s.   

In all these models the deuteron is bound largely through the tensor 
coupling between the $^3$S$_1$ and $^3$D$_1$ $N\!N$ states.
In the AV28 model, the singlet state gets a considerable part of its 
attraction through the tensor coupling between the $^1$S$_0$ $N\!N$ and 
intermediate $^5$D$_0$ $N\Delta$ states, and thus has a sensitivity 
to changes in the pion mass similar to the deuteron.
This behavior of the singlet state, i.e., that it parallels the deuteron, 
is different from that predicted by chiral perturbation theory evaluated at 
the next-to-leading-order (NLO) in Ref.~\cite{BBOS06}, which does not
include the effect of the $\Delta$.  We expect that a higher-order 
chiral perturbation calculation that includes $\Delta$ degrees of freedom
will come into qualitative agreement with our result.

For our most complete calculations, shown by the solid lines, the mass 
dependence of the energies is significantly steeper for both deuteron and
singlet states, but they all are parallel and it appears the deuteron will 
remain the ground state for a very large range of $m_q$.  The AV28 curves
shift relatively little from the OPE-only values, while the AV14, which 
is essentially phase-equivalent with AV28, gives very similar results.
The biggest change and the most-rapid dependence is given by the AV18
model, with the most important contribution coming from its static
two-pion-exchange terms, i.e., the line $m_\pi~(+TPE-s)$ in 
Table~\ref{tab:two}.  The more rapid dependence is a consequence of the
deeper intermediate-range attraction and stronger short-range repulsion,
which in turn may be a consequence of the improved quality
of AV18, i.e., that it is a better fit to more recent $N\!N$ data.

The two-body energies can also be evaluated using Eq.~(\ref{eq:energy}) to
combine the DSE values for $\delta m_H/m_H$ and the dimensionless derivatives
$\Delta\mathcal{E}(m_H)$ of Table~\ref{tab:two}.  This might not be expected
to work for large changes in $m_q$, or where a state is barely bound or
unbound.  However, deuteron energies are reproduced for changes of $\pm 0.1$ in 
$\delta m_q/m_q$ to 1\% or better by Eq.~(\ref{eq:energy}).

\begin{figure}
\includegraphics[angle=-90,width=6.25in]{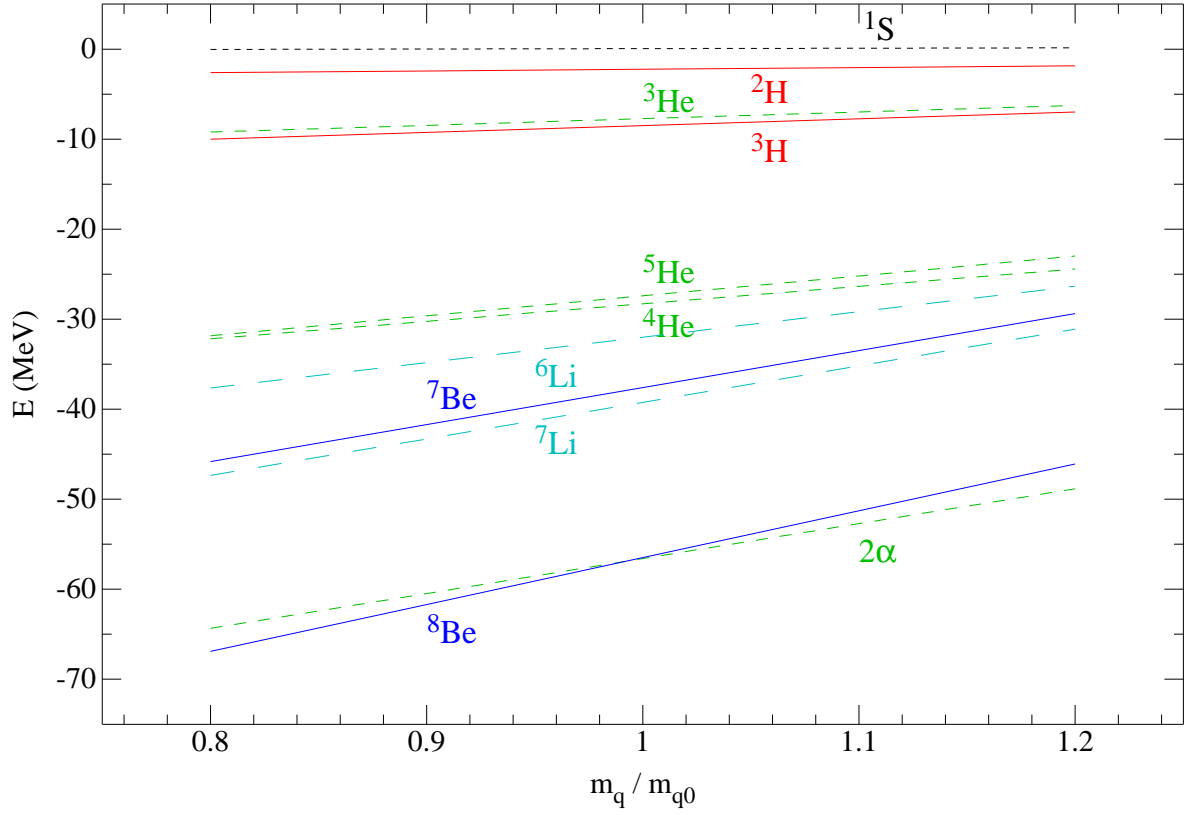}
\caption{(Color online) Variation of multi-nucleon energies with current-quark
mass variation from DSE calculation for AV14 Hamiltonian.}
\label{fig:av14}
\end{figure}

\begin{figure}
\includegraphics[angle=-90,width=6.25in]{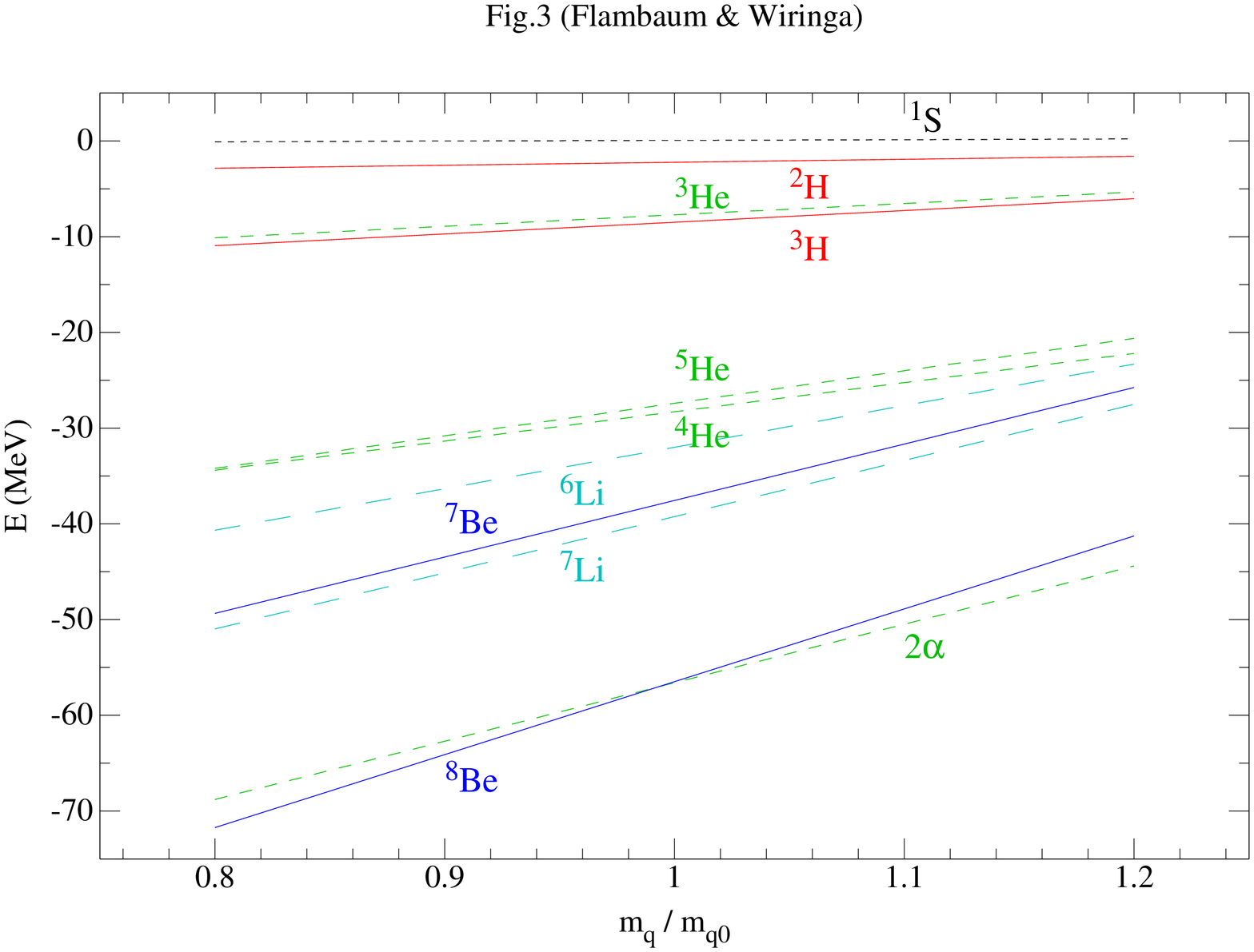}
\caption{(Color online) Variation of nuclear energies with current-quark
mass variation from DSE calculation for AV18+UIX Hamiltonian.}
\label{fig:av18uix}
\end{figure}

The dependence of the multi-nucleon energies for the full calculation with
DSE values is shown in Fig.~\ref{fig:av14} for AV14 and in 
Fig.~\ref{fig:av18uix} for AV18+UIX.  
These results have been calculated using Eq.~(\ref{eq:energy})
and the $\Delta\mathcal{E}(m_H)$ of Tables~\ref{tab:av14} and 
\ref{tab:av18uix}, respectively.  
The lines have been extended to $\delta m_q = \pm 0.2 m_q$ to show the
trends, although the results are not expected to be completely linear over 
such a broad range.  The values have been checked in a few cases by doing 
direct calculations and adjusting the variational parameters to reminimize
the energy.  For example, the $^4$He energy can be lowered by 0.3-0.5 MeV
at either end of its line, but this change would hardly be visible at the
scale shown.

The multi-nucleon energies parallel the deuteron, with generally increasing
slope as the binding energy increases.  
(Similar results for the triton were found in Refs.~\cite{EHMN06,HPP07}.)
The curves are steeper for AV18+UIX than for AV14.  
In either case it appears that the relative 
stability of all the nuclei will be preserved across a broad range of 
$m_q$ values, with the exception of $^8$Be.  Here it appears that 
a decrease in $m_q$ of $\approx$ 0.5\% will lead to $^8$Be stable against
breakup into two $\alpha$'s for both Hamiltonians. 
A very weakly bound ($\sim 0.1$ MeV) $^8$Be might not have much of an
impact on primordial nucleosynthesis because it would be easily 
photo-disintegrated until quite late in the BBN epoch.
A moderate binding ($\sim 1$ MeV) could have a significantly 
more dramatic effect on the chemical evolution of the Universe by giving 
rise to the production of noticeable amounts of stable elements with A=9,10.
If the bound state persisted to the era of star formation, it would 
presumably also have a significant effect on stellar evolution.

A summary of sensitivities of nuclear binding energies to the quark mass 
$m_q$, as given by the DSE hadronic mass variation, is presented in 
Table~\ref{tab:mq} for the different Hamiltonians.
The total sensitivity $K$,
\begin{equation}
  K = \frac{\delta E/E}{\delta m_q/m_q}
\end{equation}
of deuterium binding energy to the light quark mass is $K_d=-1.39$ 
for the AV18 interaction, while the pion contribution ranges from 
$K^{\pi}_d=-0.70$ for the OPE contribution only to $K^{\pi}_d=-3.36$
when the full TPE is counted.   This may be compared with the pion 
contribution  $3>K^{\pi}_d>-18$ from Ref.~\cite{FlambaumShuryak2002}
and $K^{\pi}_d=-2.4$ from Ref.~\cite{EMG}.

The result of Ref.~\cite{DFW04} suggested that a reduced deuteron binding
energy of $\Delta Q = -0.019 \pm 0.005$ would yield a better fit to 
observational data (the WMAP value of $\eta$ and measured $^2$H, $^4$He, and
$^7$Li abundances) for big bang nucleosynthesis.
This would correspond to an increase in the quark mass of 
$\delta X_q/X_q = 0.014\pm 0.004$ (here $X_q=m_q/\Lambda_{QCD}$).

Dent, Stern, and Wetterich~\cite{BBN} calculated the sensitivity of BBN 
abundances for $^{2}$H, $^{4}$He and $^{7}$Li to the variation of binding 
energies of $^{2,3}$H, $^{3,4}$He, $^{6,7}$Li and $^{7}$Be in a linear 
approximation.  We use the response matrix values in their Table~1
for $m_N$ to $B_{7{\rm Be}}$ and multiply by the $\delta m_N/m_N$ and 
$K$ values of our Table~\ref{tab:mq} to estimate the sensitivity of 
BBN yields to variation of the quark mass.  If we
compare to the ratio of observation and theoretically predicted
abundances given in their Appendix~B, we obtain
the following equations for $^{2}$H, $^{4}$He and $^{7}$Li:
\begin{equation}\label{H}
1 + 7.7 x = \frac{2.8 \pm 0.4}{2.61 \pm 0.04} = 1.07 \pm 0.15 \ ,
\end{equation}
\begin{equation}\label{He}
1 - 0.95 x = \frac{0.249 \pm 0.009}{0.2478 \pm 0.0002} = 1.005 \pm 0.036 \ ,
\end{equation}
\begin{equation}\label{Li}
1 - 50 x = \frac{1.5 \pm 0.5}{4.5 \pm 0.4} = 0.33 \pm 0.11 \ ,
\end{equation}
where $x=\delta X_q/X_q$.  
These equations yield 3 consistent values of $x$: 
$0.009 \pm 0.019$, $-0.005 \pm 0.038$ and $0.013 \pm 0.002$.
The statistically weighted average of $\delta X_q/X_q=0.013 \pm 0.002$ 
is dominated by the $^{7}$Li data.  A more accurate calculation
should take into account the effect of the $^8$Be binding energy variation 
(which is not calculated in Ref.~\cite{BBN}), the variation of the virtual 
$^1$S$_0(np)$ level, and non-linear corrections in $x$ which are important 
for $^{7}$Li.
Allowing for the theoretical uncertainties we should understand this
BBN result as $\delta X_q/X_q=K \cdot (0.013 \pm 0.002)$ where $K \sim 1$
and the expected accuracy in $K$ is about a factor of 2.  Note that here we
neglect effects of the strange quark mass variation.  A rough estimate
of these effects on BBN due to the deuteron binding energy variation
was made in Refs.~\cite{FlambaumShuryak2002,DFW04}.

\section{Conclusions}

We have argued that there are several reasons to question the spatial
and temporal invariance of various fundamental ``constants'' of nature, 
such as the fine structure constant $\alpha$ or a comparable strong 
interaction parameter $X_q = m_q / \Lambda_{QCD}$.
The search for evidence of such variations is ongoing in areas as diverse
as quasar absorption spectra, the Oklo natural nuclear reactor, and big bang
nucleosynthesis.

In this work we have examined how nuclear binding energies depend on hadronic
masses, including $m_N$, $m_\Delta$, $m_\pi$, and a generic heavy meson $m_V$.
We have done this by identifying the mass-dependence in several realistic 
Hamiltonians --- interactions that fit $N\!N$ elastic scattering data and 
reproduce light nuclei binding energies reasonably well in quantum Monte Carlo
calculations.  By making small changes in the masses and re-evaluating the
energy, we have obtained the dimensionless derivatives of the energy with
respect to variations in the hadronic masses.

We have combined these results with a specific prediction from a
Dyson-Schwinger equation study of sigma terms in the light-quark hadrons
for the hadronic mass-dependence on the quark mass $m_q$.  
The pion mass changes most rapidly with changes in $m_q$, so we find that 
both the one- and two-pion exchange parts of the $N\!N$ interaction are 
very important for the consequent variations in nuclear binding.
With our most complete model, the $^1$S$_0$ virtual bound state
and deuteron vary in concert; if $X_q$ increases, they both
become less bound, while if $X_q$ decreases, they both become more bound.
(We note that this result is in disagreement with chiral perturbation
results at the NLO level which have $^1$S$_0$ and $^3$S$_1$ scattering
varying antithetically.)
The binding energies of $A$=3-8 nuclei behave in the same manner, all moving
up or down together, with a sensitivity $K$ in the range $-1$ to $-1.5$.

Finally, we have folded these results with a study of the sensitivity
of big bang nucleosynthesis to variations in nuclear binding.
We find that a small increase in the quark mass of order 1\% at the time
of BBN is sufficient by itself to resolve existing discrepancies 
between theoretical and measured abundances of $^2$H, $^4$He, and $^7$Li.

\newpage

\acknowledgments

We thank C.D.\ Roberts, G.A.\ Miller, D.R.\ Phillips, and T.\ Dent 
for valuable comments.
This work is supported by the U.S.\ Department of Energy, Office of
Nuclear Physics, under contract DE-AC02-06CH11357, and by
the Australian Research Council.

\vfill
\newpage

\begin{table}[ht!]
\caption{Ground state energies of light nuclei in MeV for the different 
Hamiltonians used in this work compared to experiment.}
\begin{ruledtabular}
\begin{tabular}{lrrrrrrrrrr}
& \multicolumn{1} {c}{$^1$S$_0(np)$}
& \multicolumn{1} {c}{$^2$H}  & \multicolumn{1} {c}{$^3$H}
& \multicolumn{1} {c}{$^3$He} & \multicolumn{1} {c}{$^4$He}
& \multicolumn{1} {c}{$^5$He} & \multicolumn{1} {c}{$^6$Li}
& \multicolumn{1} {c}{$^7$Li} & \multicolumn{1} {c}{$^7$Be}
& \multicolumn{1} {c}{$^8$Be} \\
\colrule
 AV28      & 0.0661   & $-2.2250$  &         &         &          &          
&          &          \\
 AV14      & 0.0663   &  $-2.2250$ &  $-7.50$ & $-6.88$ & $-23.60$ & $-21.26$ 
& $-24.31$ & $-28.31$ & $-26.85$   & $-40.26$ \\
 AV18+UIX  & 0.0665   &  $-2.2246$ &  $-8.25$ & $-7.49$ & $-27.50$ & $-25.26$ 
& $-28.22$ & $-33.33$ & $-31.74$   & $-48.50$ \\
 Expt.     &          &  $-2.2246$ &  $-8.48$ & $-7.72$ & $-28.30$ & $-27.41$ 
& $-31.99$ & $-39.24$ & $-37.60$   & $-56.50$ \\
\end{tabular}
\end{ruledtabular}
\label{tab:energy}
\end{table}

\begin{table}[ht!]
\caption{Dimensionless derivatives
 $\Delta{\mathcal E}(m_H)
 = \frac{\delta E/E}{\delta m_H/m_H}$ of the energy
for the $^1$S$_0(np)$ virtual bound state $\epsilon_v$ and the deuteron $Q$
for all three Hamiltonians.}
\begin{ruledtabular}
\begin{tabular}{lrrrrrr}
$m_H$
& \multicolumn{3} {c}{$\Delta\epsilon_v$} & \multicolumn{3} {c}{$\Delta Q$} \\
\cline{2-4}
\cline{5-7}
& \multicolumn{1} {c}{AV28}  & \multicolumn{1} {c}{AV14}
& \multicolumn{1} {c}{AV18}
& \multicolumn{1} {c}{AV28}  & \multicolumn{1} {c}{AV14}
& \multicolumn{1} {c}{AV18}  \\
\colrule
$m_N$           & $-88.1$ & $-32.6$ & $-33.4$ & 13.06 &  8.63 &   8.90 \\
$m_N+\delta_N$  &         & $-91.2$ &$-121.2$ &       & 13.03 &  17.82 \\
$m_\Delta$      &   63.9  &         &         &$-5.15$&       &        \\
$\delta_\Delta$ &         &   68.1  &  102.2  &       &$-5.12$&$-10.36$\\
$m_\pi$ (OPE)   &    9.5  &  $-4.1$ &  $-3.8$ &$-2.23$&$-1.55$& $-1.40$\\
$m_\pi$ (+TPE-s)&   24.4  &   35.5  &   53.0  &$-3.63$&$-4.02$& $-6.70$\\
$m_\pi$ (+TPE-L)&         &         &         &$-4.02$&$-4.31$& $-6.74$\\
$m_V$           &$-153.7$ &$-245.0$ &$-381.9$ & 20.88 & 22.92 &  41.74 \\
\end{tabular}
\end{ruledtabular}
\label{tab:two}
\end{table}

\begin{table}[ht!]
\caption{$\Delta{\mathcal E}(m_H)$ for the AV14 Hamiltonian.}
\begin{ruledtabular}
\begin{tabular}{lrrrrrrrr}
& \multicolumn{1} {c}{$^3$H}
& \multicolumn{1} {c}{$^3$He} & \multicolumn{1} {c}{$^4$He}
& \multicolumn{1} {c}{$^5$He} & \multicolumn{1} {c}{$^6$Li}
& \multicolumn{1} {c}{$^7$Li} & \multicolumn{1} {c}{$^7$Be}
& \multicolumn{1} {c}{$^8$Be} \\
\colrule
$m_N$           &   6.00  &   6.44  &   3.97  &   4.58  &   5.25  &   5.60
&   5.88  &   5.10  \\
$m_N+\delta_N$  &  12.32  &  13.17  &   9.03  &  10.38  &  11.35  &  12.74
&  13.41  &  11.71  \\
$\delta_\Delta$ & $-7.35$ & $-7.82$ & $-5.89$ & $-6.74$ & $-7.10$ & $-8.31$
& $-8.76$ & $-7.69$ \\
$m_\pi$ (OPE)   & $-0.45$ & $-0.50$ & $-0.20$ & $-0.24$ & $-0.36$ & $-0.30$
& $-0.32$ & $-0.23$ \\
$m_\pi$ (+TPE-s)& $-4.35$ & $-4.66$ & $-3.33$ & $-3.87$ & $-4.19$ & $-4.83$
& $-5.09$ & $-4.38$ \\
$m_\pi$ (+TPE-L)& $-4.53$ & $-4.85$ & $-3.47$ & $-4.04$ & $-4.40$ & $-5.06$
& $-5.34$ & $-4.59$ \\
$m_V$           &  29.36  &  31.30  &  23.60  &  27.09  &  28.98  &  33.72
&  35.55  &  30.98  \\
\end{tabular}
\end{ruledtabular}
\label{tab:av14}
\end{table}

\begin{table}[ht!]
\caption{$\Delta{\mathcal E}(m_H)$ for the AV18+UIX Hamiltonian.}
\begin{ruledtabular}
\begin{tabular}{lrrrrrrrr}
& \multicolumn{1} {c}{$^3$H}
& \multicolumn{1} {c}{$^3$He} & \multicolumn{1} {c}{$^4$He}
& \multicolumn{1} {c}{$^5$He} & \multicolumn{1} {c}{$^6$Li}
& \multicolumn{1} {c}{$^7$Li} & \multicolumn{1} {c}{$^7$Be}
& \multicolumn{1} {c}{$^8$Be} \\
\colrule
$m_N$           &   6.07  &   6.54  &   3.99  &   4.51  &   5.12  &   5.24  
&   5.49  &   4.81  \\
$m_N+\delta_N$  &  16.56  &  17.73  &  11.86  &  13.31  &  14.41  &  15.53  
&  16.29  &  14.36  \\
$\delta_\Delta$ &$-12.20$ &$-13.02$ & $-9.16$ &$-10.24$ &$-10.80$ &$-11.96$
&$-12.56$ &$-11.11$ \\
$m_\pi$ (OPE)   & $-0.37$ & $-0.42$ & $-0.19$ & $-0.24$ & $-0.36$ & $-0.29$
& $-0.30$ & $-0.23$ \\
$m_\pi$ (+TPE-s)& $-6.90$ & $-7.38$ & $-5.11$ & $-5.82$ & $-6.33$ & $-6.95$
& $-7.30$ & $-6.34$ \\
$m_\pi$ (+TPE-L)& $-6.87$ & $-7.36$ & $-5.06$ & $-5.75$ & $-6.24$ & $-6.84$
& $-7.18$ & $-6.24$ \\
$m_\pi$ (+TNI)  & $-6.91$ & $-7.40$ & $-5.12$ & $-5.82$ & $-6.31$ & $-6.91$
& $-7.26$ & $-6.31$ \\
$m_V$           &  47.98  &  51.23  &  36.34  &  40.87  &  43.48  &  48.11
&  50.53  &  44.40  \\
\end{tabular}
\end{ruledtabular}
\label{tab:av18uix}
\end{table}

\begin{table}[ht!]
\caption{Dimensionless derivatives
 $K=\frac{\delta E/E}{\delta m_q/m_q}$ of the energy over
light quark mass  $m_q$ for the different Hamiltonians.}
\begin{ruledtabular}
\begin{tabular}{lrrrrrrrrrr}
& \multicolumn{1} {c}{$^1$S$_0(np)$}
& \multicolumn{1} {c}{$^2$H}  & \multicolumn{1} {c}{$^3$H}
& \multicolumn{1} {c}{$^3$He} & \multicolumn{1} {c}{$^4$He}
& \multicolumn{1} {c}{$^5$He} & \multicolumn{1} {c}{$^6$Li}
& \multicolumn{1} {c}{$^7$Li} & \multicolumn{1} {c}{$^7$Be}
& \multicolumn{1} {c}{$^8$Be} \\
\colrule
AV28     & 4.5 & -0.75 &       &       &       &       &       &        
&        &         \\
AV14     & 7.3 & -0.84 & -0.89 & -0.96 & -0.69 & -0.81 & -0.89 & -1.03  
& -1.09  &  -0.92  \\
AV18+UIX &11.4 & -1.39 & -1.44 & -1.55 & -1.08 & -1.24 & -1.36 & -1.50  
& -1.57  &  -1.35  \\
\end{tabular}
\end{ruledtabular}
\label{tab:mq}
\end{table}

\end{document}